%% file: intersection.tex
\documentclass[copyright,creativecommons]{eptcs}
\providecommand{\event}{RTRTS 2010} 
\usepackage{breakurl}             

\usepackage{graphicx}
\usepackage{color}
\usepackage{alltt}
\usepackage{shortvrb}  
\usepackage{dsfont}    

\usepackage{amsmath}
\usepackage{amsfonts}
\usepackage{amssymb}
\usepackage{subfig}

\newcommand{\attone}{\mbox{$att_1$}}
\newcommand{\attn}{\mbox{$att_n$}}
\newcommand{\sone}{\mbox{$s_1$}}
\newcommand{\sn}{\mbox{$s_n$}}

\newcommand{\hide}[1]{}

\title{Specification and Verification of Distributed Embedded Systems: 
A Traffic Intersection Product Family}
\author{Peter Csaba {\"O}lveczky
\institute{Department of Informatics \\  University of Oslo}
\and
Jos\'e Meseguer
\institute{Department of Computer Science \\ University of Illinois at Urbana-Champaign}
}
\def\titlerunning{Specification and Verification of  a Traffic Intersection Product Family}
\def\authorrunning{P. C. {\"O}lveczky and J. Meseguer}
\begin{document}
\maketitle

\begin{abstract}
\input{abstract}
\end{abstract}

\input{intro}

\MakeShortVerb{\@}    
\input{real-time-maude}
\input{req}

\input{overview}

\input{concl}

\bibliographystyle{eptcs} 

\input{intersection.bbl}
\end{document}

%% file: abstract.tex
Distributed embedded systems (DESs) are no longer the
  exception; they are the rule in many application areas such as
  avionics, the automotive industry, traffic systems, sensor networks,
  and medical devices.  Formal DES specification and verification is
  challenging due to state space explosion and the need to support
  real-time features.  This paper reports on an extensive
  industry-based case study involving a DES product family for a
  pedestrian and car 4-way traffic intersection in which autonomous
  devices communicate by asynchronous message passing without a
  centralized controller.  All the safety requirements and a liveness
  requirement informally specified in the requirements document have
  been formally verified using Real-Time Maude and its model checking
  features. 

%% file: intro.tex
\section{Introduction}

Distributed embedded systems (DESs) are no longer the exception; they
are the rule in many application areas such as avionics, the
automotive industry, traffic systems, sensor networks, and medical
devices.  The specification and verification of such systems poses
special challenges: besides modeling the real-time behavior of each
distributed component, faithfully modeling the real-time 
\emph{interactions} between the different embedded components is just
as crucial.  This is hard because the asynchronous nature of component
interactions makes verification particularly challenging due to state
space explosion. 

Furthermore, in an industrial setting there are additional reusability
requirements for DES specification and verification: (1) the different
components of the DES should be specified in as \emph{modular and
  reusable} a way as possible; and (2) rather than focusing on a
single DES design, the specification and verification effort should be
amortized over an entire \emph{parametric family of DES designs},  which
can be used to develop \emph{product families} in order 
to cut development cost
and shorten time to market without diminishing system quality or
compromising safety.  This means
that component specifications should be both modular and parametric,
making it easy to obtain different product designs by composing
different components from a library and fixing the various
parameters of each component. 

This paper reports on an extensive case study about a DES parametric
design that can serve as a basis for a product
family for a pedestrian and car 4-way traffic intersection that should
be totally decentralized (no central controller), should support
different traffic light protocols such as the American and European
light regimes, should be strongly fault-tolerant, and should have
sophisticated features such as detecting the proximity of emergency
vehicles and then switching the entire distributed system to an emergency
regime in which emergency vehicles have absolute priority over
ordinary vehicles and pedestrians. 

One important feature of this case study is that the requirements that
the specification and verification of this product family should meet
were not determined by us: they were provided by an industrial partner
(Lockheed-Martin) with no input on our part and with a substantial
degree of realism.  Furthermore, the traffic intersection product
family was viewed as a representative example exhibiting many of the
challenges that would be present in other DES product families such
as, for example, command and control systems.  Specifically, the
informal system requirements were specified as a Statement of Work
(SoW) that should be abided by a system design team. The SoW included
a number of important safety 
and liveness requirements that the system design should meet.

The overall goal of this case study was to evaluate on a realistic and
somewhat challenging application the suitability of different modeling
languages for the design and analysis of DES product families.  Since
many system aspects needed to be addressed, including deployment of
the DES design on specific hardware platforms, system cost estimation,
performance evaluation, and system safety, no single modeling language
was envisioned.  Instead, the emphasis was 
on \emph{multimodeling}~\cite{naomi-multi-model},
with different modeling languages handling those aspects that they 
do best. Real-Time Maude was chosen to handle all formal modeling and verification
aspects, including the verification of all safety, fault-tolerance, and liveness requirements.

\subsection{Challenges, Experience, and Contributions}

We describe here the formal modeling and verification challenges that had to be addressed
in order for Real-Time Maude to successfully handle this case study.  For each such challenge
we explain how we addressed it.  We then finish giving a summary of our overall experience
and our contributions.

\paragraph{Modular, Decentralized, and Asynchronous Design.}
To address this challenge we model the traffic light system as a
collection of autonomous concurrent objects that interact with each
other by asynchronous message passing. This makes the system highly
modular and reusable. In particular, the system has no central
controller; instead, the different components interact with each other
through messages and act autonomously.  The system is also highly
parametric: ten different parameters can be specified to obtain
different product versions with different special features, such
as, for example, support for American or European light regimes, and reaction to the presence of
emergency vehicles in the environment.  Fault
tolerance and recovery issues are explicitly addressed: failures are
modeled explicitly under a very general fault model in which
multiple devices can fail at any time. 

\paragraph{State Space Explosion.}
It is well-known that concurrency and asynchrony can easily cause an
exponential
 blowup in the search space of a system when model checking its properties. 
Because of the highly concurrent and asynchronous nature of our traffic light 
design this poses an important modeling and verification challenge.
We have addressed this challenge without sacrificing the good features
of modularity, decentralization, autonomy, and asynchrony in the
design. It would of  
course have been much easier to model check a highly centralized synchronous 
design with much fewer states. But this would have meant sacrificing many good 
design aspects for the sole purpose of making our verification task easier. Also 
this would have rendered the design less reusable and would have considerably 
widened the gap between the model and a realistic deployment. 
The key approach that we have taken to achieve a design that can be
 feasibly verified without compromising good design features has been the use of 
\emph{abstraction}. This means that the model is \emph{as abstract as
possible, yet, all 
relevant features are modeled}. For example, since in this example, in the north-south (N-S) and 
south-north (S-N) directions (similarly for east-west and west-east)
the car 
lights will always have the 
exact same color, it is not necessary to have two light devices (one for N-S-bound 
cars, and another for S-N-bound cars). Instead, it is enough to have a single light 
device for the two (N-S and S-N) directions, in the implicit understanding that 
such a device will have two simultaneous and identical light displays for both the 
N-S and S-N directions. In a similar vein, the message-passing communication 
between devices is assumed to be instantaneous, abstracting out the implicit 
assumption of a tolerable bound in the network communication delays, given 
that light changes happen at the level of seconds, whereas network delays can 
be assumed to be in the order of milliseconds. 
However, we believe that all relevant aspects of the system are still modeled, 
including failures and behavior for emergency vehicles. The only aspects not 
modeled are those needed for performance estimation purposes, which are not 
relevant for safety purposes, and that do not preclude certain, more abstract 
analyses of liveness properties. Specifically, we do not model the exact number 
of cars or pedestrians, but only their presence or absence near the intersection. 

This does not exclude the possibility of developing much more detailed 
models of this system in Real-Time Maude for simulation and performance estimation 
purposes. This could easily be done; but their direct formal verification by model 
checking would be unfeasible. However, they could also be proved correct 
indirectly, by proving that the model that we present here is a correct abstraction 
of these more detailed models.

\paragraph{Formal Verification of Safety and Liveness Properties.}
As we explain in more detail in the body of the paper, all the safety
properties mentioned in the SoW, as well as a liveness property,
have been verified. This means that precise formal specifications in the
form of temporal logic formulas have been developed for each of the
informal and somewhat vague corresponding requirements in the SoW
document; and then they have been verified by model checking in our
model. It also means that, as we further explain later, certain
inconsistencies between the informal requirements have also been
identified and addressed in both the model design and the formal
verification. 

\paragraph{Overall Experience and Main Contributions.}
Our overall experience can be summarized as follows.  Because of its
support for distributed objects and asynchronous communication, we
have been able to effectively use Real-Time Maude to develop a highly
reusable and modular distributed system design for the 4-way traffic
intersection system in the form of a parametric product family that
could be instantiated in various ways to support a variety of
additional features.  Perhaps the hardest challenge has been to avoid
state space explosion without compromising the faithfulness with which
the formal model captures all relevant system aspects.  Here, careful
use of abstraction to exclude all non-essential aspects from the model
has been crucial.  This made it possible for us to formally verify all
informal safety requirements in the SoW, plus one important liveness
requirement.  The overall experience has been quite positive, in that
all the modeling challenges posed by the case study could be
successfully addressed and all the expected verification tasks were
accomplished. 

We are not aware of other industrially-based case studies in which
full formal specification and verification of DES systems with
comparable degrees of concurrency, modularity, and parametricity have
been carried out.  In this regard we view our main contribution to be
a novel and convincing demonstration that formal specification and
verification of nontrivial DES product families is indeed possible in
spite of the challenges involved; and that rewriting-based methods
such as those supported by Real-Time Maude are indeed effective in
addressing these challenges.  A subsidiary contribution is the use of
a distributed object-based formal modeling to achieve a 
decentralized and highly reusable DES product family design.

%% file: real-time-maude.tex
\section{Real-Time Maude}
\label{sec:rtm}

A Real-Time Maude \emph{timed module} specifies a  \emph{real-time
rewrite theory}  of the form
 $(\Sigma, E, \mathit{IR}, \mathit{TR})$, where:
\begin{itemize}
\item  
$(\Sigma, E)$ is a \emph{membership equational
logic}~\cite{maude-book} 
theory with $\Sigma$ a signature\footnote{That is, $\Sigma$ is a set
  of declarations of \emph{sorts}, \emph{subsorts}, and
  \emph{function symbols}.} and $E$ a set of {\em confluent and terminating
conditional equations}. 
$(\Sigma, E)$
specifies
the system's state space as an algebraic data type, and must 
  contain a specification of a sort @Time@ modeling the (discrete or dense)
time
domain. 

\item  $\mathit{IR}$
is a set of (possibly conditional) \emph{labeled instantaneous rewrite
  rules} specifying  
the system's \emph{instantaneous} (i.e., zero-time) local transitions,  written 
$@rl [@l@] : @t@ => @t'$,
where  $l$ is a \emph{label}. 
 The rules are applied \emph{modulo} the
equations~$E$.\footnote{$E = E'\cup
  A$, where $A$ is a set of equational axioms such as associativity,
  commutativity, and identity, so that deduction is performed \emph{modulo}
  $A$. Operationally,  a
term is reduced to its
$E'$-normal form modulo $A$ before any rewrite rule is applied.}

\item $\mathit{TR}$ is a set of \emph{tick (rewrite) rules}, written
  \quad  \texttt{rl [\(l\)]\!\! :\!\!\! \char123\(t\)\char125{}  => \char123\(t'\)\char125{} in time \(\tau\)},\;
that model time elapse.  @{_}@ is a 
built-in
 constructor of  sort \texttt{GlobalSystem}, and
$\tau$ is a term of sort @Time@ that denotes the \emph{duration}
of the rewrite.
\end{itemize}
The initial state must be a ground term of sort @GlobalSystem@ and 
 must be reducible to a term of
the form @{@$t$@}@ using the equations in the specifications. 

The Real-Time Maude syntax  is fairly intuitive. For example, 
 function symbols, or \emph{operators}, 
 are declared with the syntax \texttt{op }$f$ @:@ $s_1$ \ldots $s_n$
 @->@ $s$. $f$ is the name of the
 operator;
$s_1\:\ldots\:s_n$ are the sorts of the arguments of $f$; and $s$ 
 is its (value) \emph{sort}. Equations are written
with syntax @eq@ $t$ @=@ $t'$, and @ceq@ $t$ @=@ $t'$ @if@ \emph{cond}
for conditional equations. The mathematical variables in such statements
are declared with the keywords {\tt var} and {\tt vars}.
We refer to~\cite{maude-book} for more details  on the syntax of
 Real-Time Maude.

In object-oriented Real-Time Maude modules, a \emph{class} declaration

\small
\begin{alltt}
  class \(C\) | \(\attone\) : \(\sone\), \dots , \(\attn\) : \(\sn\) .
\end{alltt}
\normalsize

\noindent declares a class $C$ with attributes $att_1$ to $att_n$ of
sorts $s_1$ 
to $s_n$, respectively. An {\em object\/} of class $C$ in a  given state is
represented as a term
$@<@\: O : C \mid att_1: val_1, ... , att_n: val_n\:@>@$
of sort @Object@, where $O$, of sort @Oid@,  is the
object's
\emph{identifier}, and where $val_1$ to 
$val_n$ are the current values of the attributes $att_1$ to
$att_n$, respectively.
 In a concurrent object-oriented
system, the 
 state
 is a term of 
the sort @Configuration@. It  has 
the structure of a  \emph{multiset} made up of objects and messages.
Multiset union for configurations is denoted by a juxtaposition
operator (empty
syntax) that is declared associative and commutative, so that rewriting is 
\emph{multiset
rewriting} supported directly in Real-Time Maude.
The dynamic behavior of concurrent
object systems is axiomatized by specifying its 
transition patterns by  rewrite rules. For example, 
  the rule

{\small
\begin{alltt}
rl [l] : m(O,w)  < O : C | a1 : x, a2 : O', a3 : z >   =>
                 < O : C | a1 : x + w, a2 : O', a3 : z >  m'(O',x) .
\end{alltt}
}

\noindent  defines a  family of transitions 
in which a message @m@, with parameters @O@ and @w@, is read and
consumed by an object @O@ of class @C@. The transitions have the 
 effect of altering
the attribute @a1@ of the  object @O@ and of sending a new message
@m'(O',x)@.  
``Irrelevant'' attributes (such as @a3@)
need not be mentioned in a rule.

\paragraph{Formal Analysis.} 
A Real-Time Maude specification is \emph{executable}, and the tool offers a variety of formal analysis 
methods. In this paper we focus on temporal logic model checking. 
Real-Time Maude  extends Maude's \emph{linear temporal logic model
  checker} 
  to check whether
each behavior, possibly  up to a certain time bound,
  satisfies a temporal logic 
  formula.
 \emph{State propositions} are terms of sort @Prop@, and their 
semantics is defined by (possibly conditional) equations of the form
\texttt{\char123\(\mathit{statePattern}\)\char125{} |= \(\mathit{prop}\) = \(b\)}, 
with $b$ a term of sort @Bool@.  Such equations define the state proposition 
$prop$ to hold in exactly those
states $@{@t@}@$ where $@{@t@}@$ \verb+|=+ $prop$ evaluates to @true@.
A temporal logic \emph{formula} is constructed by state
propositions and
temporal logic operators such as @True@, @False@, @~@ (negation),
@/\@ (conjunction), @->@ (implication), @[]@ (``always''), @<>@
(``eventually''), and @U@ (``until'').
The  model checking command is written \quad
 \texttt{(mc \(t\) |=u \(\mathit{formula}\) .)}\quad 
for $t$ the initial state and $\mathit{formula}$  the  temporal logic formula.

Finally, Real-Time Maude provides some \emph{metric} temporal logic model
checking comands for non-hierarchical object-oriented models, such as the 
traffic intersection. For example, the \emph{bounded response} model checking command
\quad \texttt{(br \(t\) |= \(p\) => <>le( \(\tau\) ) \(q\) .)}\quad 
 investigates whether each $p$-state will be followed by a $q$-state 
in $\tau$ time units or less~\cite{rtrts10-br}.

%% file: req.tex
\section{Overview of the Requirements Specification}

This section gives a brief overview of selected parts, namely,  those
concerning \emph{functionality} rather than \emph{performance},  of  the 
\emph{statement of work} (SoW) for the \emph{Easily Deployable Traffic Congestion Management System
for Four-Way Intersections} (EDeTCMS-4) that was provided to us by Lockheed Martin. 

The overall goal is to design a four-way traffic intersection solution
that is easily customizable to enable rapid deployment 
in  the US and Europe. The aim of the system is to reduce commute times and
traffic jams, but \emph{not at the expense of safety and reliability aspects which keep
motorists and pedestrians safe on the road}. In particular, a main goal is to reduce the emergency 
vehicle commute times which have become more and more unacceptable with the rising 
number of vehicles on the road.  

Some assumptions about the intersections where EDeTCMS-4 will be deployed are: (i) pedestrians walk in all possible directions \emph{except} diagonally through the intersection; (ii) traffic enters the intersection from any possible direction, and exits in either a left, straight, or right direction; (iii) emergency vehicles may require the intersection to be cleared at any time, allowing them to enter from any direction and exit in either a left, straight, or right direction; and (iv) there are no cross-traffic dedicated turn lanes.

The requirements on the light operations include:
(v) the system shall turn green a pedestrian  light only when
 pedestrians are waiting to cross in that direction; 
(vi) the system should turn green a vehicular light only when there
are vehicles waiting to go in the direction controlled by the light,
or when turning the light green does not prohibit any other cars
 from proceeding through the
intersection; (vii) the system must be fault-tolerant, and, except for transient faults,
must ensure
the proper functioning of the lights;  (viii) the system must
ensure failure recovery and safe car and pedestrian conditions also
under failures; (ix)  under no circumstances (including in an emergency clearance)
  will there be
unsafe situations for cars or pedestrians; and (x)  
the maximum pedestrian wait time shall be less than 5 minutes.

%% file: overview.tex
\section{The Real-Time Maude Model of the EDeTCMS-4}
This section presents our Real-Time Maude model of the
EDeTCMS-4 intersection. The entire executable model, including the 
model checking commands is Section~\ref{sec:analysis}, is available at 
\url{http://www.ifi.uio.no/RealTimeMaude/TrafficLight}.

\subsection{Overview and Assumptions}

We have defined a model of the behavior of the traffic lights that
should function correctly under the following 
very general conditions:
\begin{itemize}
\item 
Emergencies may happen at \emph{any} time, and
may end at any time thereafter.
\item Any  device may fail at any time. We
 assume that there is a \emph{minimum} time interval between
the repair of a device and the next time \emph{that} device can fail. This
minimum interval is a parameter of the system, and can be set to 1 to get a
completely nondeterministic failure model.
\end{itemize}

Our model focuses on \emph{modularity} and \emph{autonomy},
 in the sense that each
traffic light should  operate as independently as possible. In our model,
 devices only communicate  through message passing. 

We have defined an object-based model of EDeTCMS-4. Each intersection
has four objects modeling the traffic lights:
\begin{enumerate}
\item One object models the (controller for the) car lights in the east-west
direction.
\item One object models the (controller for the) car lights in the 
north-south direction.
\item One object models the (controller for the) pedestrian lights
in the east-west direction.
\item One object models the (controller for the) pedestrian lights
in the north-south direction.
\end{enumerate}

Since there are no turn signals, we did not see any reason why
the car lights in opposite directions (say, north and south)
should not always show the same color.  From this, it follows that both
pedestrian crossings in the same direction should have the same color.
This is also how we have observed traffic lights; e.g., you push the
button on one pedestrian light pole, and all four buttons for that
direction are lit. Therefore, we assume \emph{one} controller for the car
lights in the north/south direction, and so on. 

In addition to the controllers, we model the environment as follows:
\begin{itemize}
\item One environment object  nondeterministically generates
new cars and new pedestrians at each time instant. That is, at each
moment in time, this environment object 
 may or may not generate a new pedestrian/car in a
certain direction. 
\item One environment object generates \emph{emergency} and
\emph{emergency-over} signals. Such signals can be generated at any
time, with the \emph{emergency-over} signal following at any time
after one or more \emph{emergency} signals. 
\item For \emph{each} device/controller, 
there is a corresponding environment object
which generates failures for that device at any time. After a failure,
it may generate a repair message for the device at \emph{any time}
after the failure.  Such a repair may then be followed by a new
failure at any time after a minimum separation between a repair and a new
failure. 
\end{itemize}

By including/excluding certain of the above environment
objects in the initial state of the system,  the system can be analyzed 
in the presence/absence of the corresponding emergencies/failures. 
Section~\ref{sec:analysis}  illustrates this feature by analyzing the system without
emergencies and failures, with emergencies and without failures, with
the failures of a subset of the devices, and so on.

\subsubsection{Parametricity and Reusability}

The requirements place great emphasis on being able to use the controller in different 
intersections in 
both Europe and the U.S. Our model supports substantial 
reuse in the following ways:
\begin{itemize}
\item It is defined for both American and European traffic light
configurations. A parameter can be set to denote European or American
deployment. 
\item It is parametric in important parameters such as: 
\begin{itemize}
\item the amount of time during which the light in a given direction should be
green/red in a round,
\item the safety margin during which the car lights in both
directions should be red,
\item the time during which the light should be yellow in each round,
\item the minimum time  it is assumed to take  for a pedestrian to
cross the street, and so on.
\end{itemize}
This allows the system to be deployed
under varying circumstances. For instance, we have noticed as lone
walkers in Cherry Hill, NJ, that the system assumes that pedestrians
cross the street faster than an  Olympic sprinter,
whereas in Europe people  typically have more time to cross.
 Likewise, as the speed limit increases, so should the safety
margin and the time the light stays yellow. 
\item Since our model is designed to work under very
 general failure and emergency assumptions, it can be deployed
 in all kinds of places, including in poorer 
communities where devices often fail  and where repair is rarely
available, as well as in places where the
members of the plutocracy  activate the emergency clearance with a
high frequency and long durations.  
\item Our object-oriented model is also parametric in the number of 
intersections, so that we can deploy and  analyze a \emph{set} of
intersections by changing the initial state. 
\end{itemize}

\subsubsection{Inconsistent Requirements}

During
the formalization effort we discovered that the requirements (vi) 
(a car light should turn green only when there are cars waiting) and
(x)  (no pedestrian should wait for more than five minutes to cross)
 seem be in contradiction with each other: if no car is driving  in a
certain direction, then the cars coming from the other direction should
not be prohibited from proceeding through the intersection, so the poor
pedestrian has to wait forever.
We therefore model a system where a car light can turn green if either
cars \emph{or pedestrians} are waiting to cross in that direction.

\subsubsection{Overview of Car Lights in Normal Operation}

In a setting without failures, the car lights 
operate almost independently. The only communication between the two
car light (controllers) happens when one car light is about to turn
green, but detects that no car or pedestrian is waiting to cross in
that direction. This car light must then send a message to the other car light, 
informing that other car light that it can stay green for another round. 

The parameters for the car lights are: 
@yellowTime@ (the time that the light
is yellow after being green and before turning red), @safetyMargin@ (the
time during which both car lights should be red), and
@redTime@$_{NS}$ and @greenTime@$_{NS}$ (the duration of the red,
resp.\ green, light in the NS direction in each round).
The operation of a car light controller in direction $D$ 
can be summarized as follows:
\begin{enumerate}
\item Assume that the car light for direction $D$ just got red. 
\item It  stays  red for time $@redTime@_{D} - (\Delta +
@yellowTime@ + @safetyMargin@)$, for some small value $\Delta>0$. 
\item It then checks the sensors to see if there is a car present at the
approach, and also checks whether it has recorded a signal from the
associated pedestrian light about the pedestrian button having been
pushed lately. If a car is waiting or a pedestrian button 
 push has been recorded,
 the car light  shows red for additional time $(\Delta +
@yellowTime@ + @safetyMargin@)$, unless it is a European light, in
which case it waits for only time $\Delta + @yellowTime@$ before
turning both yellow and red. If neither car nor pedestrian is waiting,
then  the car light sends a signal to the opposing car light that it
will  not turn green this time, and resets its timer to an entire
round, and remains red.  
\item After waiting for the duration given in item (3), the red light turns
green. If the car light has recorded a pedestrian push lately,
 it
sends a message to ``its'' pedestrian light to turn  the light 
green.
\item After  showing  green for time @greenTime@$_D$, the light
turns yellow \emph{unless} it has received a signal that the other car
light does not need to turn green, in which case it stays green for another
 round. 
\item After  showing yellow for time @yellowTime@, it turns red,
and starts from item (1) above.
\item In addition, the car light treats 'pedestrian pushed button'
messages from the pedestrian light. 
\end{enumerate}
Given  @redTime@$_{NS}$ and @greenTime@$_{NS}$, the
corresponding values for the $EW$ direction 
can be computed. 

\subsubsection{Pedestrian Lights During Normal Operation} 

The operation of a pedestrian light is fairly simple:
\begin{itemize}
\item 
A pedestrian light
 turns green when it receives a 'pedestrian go' message from its car
light.
\item When time @walkTime@ remains of the time period
for which the car light has promised to stay green, the pedestrian
light starts blinking.
\item After the light has been blinking for time @walkTime@, it turns
red. 
\item When a new pedestrian arrives, and the light is red or blinking,
the pedestrian is assumed to push the button on the pedestrian light
pole unless this is lit. If the button is not lit, it becomes lit,
and a message is sent to the car light. 
\end{itemize}

\subsubsection{Emergency Clearance}

In the literature about the treatment of emergencies in
traffic systems, there  are different  ways of signaling and sensing emergencies, such as acoustic
sensing, and there are different  ways of responding to emergencies, one
of which is to turn all lights  red, another which is to turn
some light green.

Our model  has been influenced by thinking about acoustic sensing,
in that we assume that one signal is sent when the emergency situation first 
appears  (the car lights detect the sirens), and one is sent when
the emergency is over (it suddenly becomes quiet again). Furthermore,
in an acoustic setting we did not find it natural to assume that only one
car light detects the emergency signals. 

Given that there is
 a real problem of people trying to fool the sensors when the light
turns green during emergency, we have opted for what we think is the
most natural solution: all lights turn red during emergencies. 
Since drivers hear the emergency signals, they will hopefully  make way
for emergency vehicles even if that implies violating red lights if
needed. 
However, in order 
not to add  collisions when emergency vehicles arrive,
we think it is natural to turn a green light yellow before turning
it red during emergencies.
 When the emergency is over, the system restarts from a
standard starting state, which basically means that the 
\emph{prioritized direction} gets the green light first, \emph{if} there
are cars or pedestrians waiting; otherwise the direction with lowest
priority restarts with green lights.

\subsubsection{Failures}

Each device (for which an error generator is included in the initial
state of the system) may fail at any time, and will be repaired after
an arbitrary amount of time. Upon detecting a failure in a device, the
failed device notifies the other devices about the failure, and all
devices go into error mode as follows:
\begin{itemize}
\item The car lights in the prioritized direction start blinking
yellow.
\item The car lights in the other direction start blinking red. 
\item All pedestrian lights are turned off. 
\end{itemize}

Each device must keep track of how many devices are currently in
failed state, so
that they only go to normal mode after \emph{all} the failed devices have been
repaired. When all devices have been repaired, the system restarts
from a ``neutral'' position.

\subsubsection{Communication}

Given the short distances between the devices and the relatively large
time scale of the changes in the traffic lights, we abstract from
communication delays and assume instantaneous and reliable
asynchronous message
passing communication between the devices, and from the ``environment'' to the
devices.

\subsection{The Real-Time Maude Model}

This section  presents fragments of our Real-Time Maude model of
the controllers.

\paragraph{Tunable Parameters.}

The following defines some system parameters. 
 Additional parameters, such as @redTime@ and
@greenTime@, are given as parameters to the initial state as shown in Section~\ref{sec:initial-states}.

\small
\begin{alltt}
  \emph{--- American or European crossing?}
  ops americanXing europeanXing : -> Bool .  
  eq americanXing = true .          eq europeanXing = false . 
  \emph{--- A small amount of time:}
  op Delta : -> NzTime .            eq Delta = 1 . 
  \emph{--- Safety margin is the time that both lights should be red:}
  op safetyMargin : -> NzTime .     eq safetyMargin = 1 .   
  \emph{--- Duration of yellow light before turning red:}
  op yellowTime : -> NzTime .       eq yellowTime = 1 .        
  \emph{--- The shortest time it takes to cross the street for a pedestrian:}
  op walkTime : -> NzTime .         eq walkTime = 2 .      
  \emph{--- Minimum duration of green and red car lights that ensures that}
  \emph{--- pedestrian will see some green before blinking:} 
  ops minRedTime minGreenTime : -> NzTime .
  eq minGreenTime = walkTime + 1 .
  eq minRedTime = safetyMargin + minGreenTime + yellowTime + safetyMargin .
\end{alltt}
\normalsize

\paragraph{Object Identifiers.}

We envision that cities will have multiple
intersections, and our model supports multiple (independent)
intersections. Therefore, each object's name includes the name of the
intersection. For example, the name of a car light controller could be 
@carLight("SpitsB-2", NS)@, denoting the car lights for the north-south
direction in the intersection called @"SpitsB-2"@. The corresponding
pedestrian light controller object should be named
 \texttt{pedLight("SpitsB-2", NS)}. Furthermore, 
@approach(@$\mathit{xing},\: \mathit{dir}$@)@ is the name of
the sensor which senses whether some car is traveling in the direction
$\mathit{dir}$ in the intersection $\mathit{xing}$; and so on:

\small
\begin{alltt}
 vars CN CN' : CrossingName .   var DIR : Direction .
 ops NS EW : -> Direction [ctor] .
 op opposite\! :\! Direction -> Direction\! .\!   eq opposite(NS) = EW\! .\!   eq opposite(EW) = NS\! .

 op carLight : CrossingName Direction -> CarLightId [ctor] .
 op pedLight : CrossingName Direction -> PedLightId [ctor] .

 ops cl pl : DeviceId -> DeviceId .     
 eq cl(pedLight(CN, DIR)) = carLight(CN, DIR) .
 eq pl(carLight(CN, DIR)) = pedLight(CN, DIR) .

 op opposite : DeviceId -> DeviceId .
 eq opposite(carLight(CN, DIR)) = carLight(CN, opposite(DIR)) .
 eq opposite(pedLight(CN, DIR)) = pedLight(CN, opposite(DIR)) .

 op approach : CrossingName Direction -> CarApproachId [ctor] .
 op pedStop : CrossingName Direction -> PedId [ctor] .
\end{alltt}
\normalsize

\paragraph{Messages.}

The following messages define the interface between the objects:
a @continueGreen(@$carL$@)@ message signals to the receiving car light 
$carL$ that it should remain green for another round; 
the @pedGo@ message signals to the receiving pedestrian light that it
should turn green and remain green/blinking for the given amount of
time;  and the @pedsWaiting@ message is sent from a pedestrian light to the
corresponding car light to signal that a pedestrian has pushed the
unlit button. 

@newPed(@$\mathit{pedLight}$@)@
and @newCars(approach(@$\mathit{xing}, \mathit{dir}$@))@ messages are generated by the environment
to denote that new (non-empty sets of) pedestrians and cars have
arrived at the given place.

 @emergency(@$\mathit{xing}$@)@ and @emergencyOver(@$\mathit{xing}$@)@
messages are generated by the environment and signal,
respectively, the start of a period of emergency and the end of such a
period. Given that we assume  acoustic sensing of sirens, we assume
that  both car lights hear any emergency
signal. Therefore, emergency messages are
 only sent to the crossing;  the two equations below then 
``distribute'' such a message to each of the two car lights in the
crossing:

\small \begin{alltt}
eq emergency(CN) = emergency(carLight(CN, EW))  emergency(carLight(CN, NS)) .
eq emergencyOver(CN) = emergencyOver(carLight(CN,\!\! EW))\!  emergencyOver(carLight(CN,\!\! NS))\!\!\! .
\end{alltt} \normalsize

\noindent The following messages are sent from a car light to the
corresponding pedestrian light to signal, respectively, that an
emergency is detected; and when the emergency is over, whether to turn
red or green.

\small \begin{alltt}
msg emergency resumeRed : PedLightId -> Msg .
msg resumeGreen : PedLightId NzTime -> Msg .
\end{alltt} \normalsize

\noindent When an emergency is over, the car lights need to restart.
 In our setting,
one of the directions is defined to be the prioritized direction which
should turn green first after an emergency. However, given requirement
(vi) in the SoW, this prioritized  light cannot turn green if no car or
pedestrian  is
waiting in that direction. Therefore, the main car light must signal
to the other car light how to restart after the emergency:

\small \begin{alltt}
msgs reStartRed reStartGreen : CarLightId -> Msg .
\end{alltt} \normalsize

\noindent Finally, the following messages are used for failure generation and
repairs. The environment generates error messages of the form 
@(to @$d$@ error(@$d$@))@. Once device $d$ reads such an error
message, it must also inform the three other controllers about its
failure by sending a @(to @$d_k$@ error(@$d$@))@ message to each other
device $d_k$ in the intersection. The messages for repairs work in the
same way. 

\subsubsection{Car Lights During  Normal Operation}

We first present the behavior of the car
lights during normal operation.  
Car light controllers are modeled as object instances  of a   class @CarLight@  with the following attributes:
\begin{itemize}
\item @lights@  denotes the
lights shown (in both directions) by the car light. It is a \emph{set} of
colors, because in Europe it may show \emph{both} red and yellow at the
same time.
\item The @timer@ attribute is set to expire when the 
next time-triggered action must be taken. 
\item @state@ denotes the ``internal'' state of the objects, as seen
in the rules below.
\item @redTime@ and @greenTime@ denote the length of time the light
stays red,  respectively green,  in a round. Since these are class
attributes and hence represented in the state, they can be changed
during system execution, for example to increase the  @greenTime@ of
the busier direction.
\item @pedWaiting@ denotes whether or not the car light has received a
signal from the pedestrian light that a pedestrian has pushed the
button. This is needed to avoid  turning a pedestrian light
green if no pedestrians are waiting in that direction. 
\item @defaultStarter@ is @true@ if this car light has priority to
turn green after failures and emergencies. 
\end{itemize}

\small \begin{alltt}
class CarLight | lights\! :\! ColorSet,\!  timer\! :\! TimeInf,\!  state : CLState, redTime\! :\! NzTime,
                 greenTime\! :\! NzTime,\!  pedWaiting\! :\! Bool,\!  defaultStarter\! :\! Bool\! .   

sorts CLState NormalCLState .    subsort NormalCLState < CLState .
ops red toRedYellow toGreen green yellow : -> NormalCLState [ctor] . 
op emergency : -> CLState [ctor] .
\end{alltt} 
\normalsize  

The following rules apply when the timer expires and the object is 
 in state @red@. The car light must then  check
 whether some cars are waiting in the
corresponding approach, or whether it has any record of a pedestrian
having pushed the button lately. If neither car nor pedestrian is
waiting, rule @dontGoGreen@ can be applied, and the car light stays in
state and color @red@ but resets its timer to expire an entire round
later ($@redTime@ + @greenTime@ + @yellowTime@$). It also sends
a @continueGreen@ message to the other car light. If, on the other
hand, some pedestrian  or  car is
waiting (@B1 or B2@),
 rule @redToSafetyMargin@ is applied. Depending on the
type of crossing, the timer is set to expire either when the light
should turn green, or when it should turn red and yellow. The light
itself is not updated, and stays red. Only the ``internal'' state and
the timer are updated. We first declare the variables used in the rules:

\small \begin{alltt}
var C\! :\! Color .         var CL\! :\! CarLightId .      var PL\! :\! PedLightId .
vars RT GT\! :\! NzTime .   vars B B1 B2\! :\! Bool .      var TI\! :\! TimeInf .
var T\! :\! Time .          var DIR\! :\! Direction .      var CN\! :\! CrossingName .
var S\! :\! CLState .       var NORMAL\! :\! NormalCLState . 

rl [dontGoGreen] :
   < approach(CN,\! DIR)\! :\! XingApproach | carsPresent\! :\! false >
   < carLight(CN,\! DIR)\! :\! CarLight | state\! :\! \emph{red},  timer\! :\! \emph{0},  pedWaiting\! :\! false, 
                         \!\!\!           redTime\! :\! RT,  greenTime\! :\! GT >
  =>
   < carLight(CN,\! DIR)\! :\! CarLight | timer\! :\! GT + RT + yellowTime >
   < approach(CN,\! DIR)\! :\! XingApproach | >
   continueGreen(carLight(CN,\! opposite(DIR))) .
  
crl [redToSafetyMargin] :
    < approach(CN,\! DIR)\! :\! XingApproach | carsPresent\! :\! B1 > 
    < carLight(CN,\! DIR)\! :\! CarLight | state\! :\! \emph{red},  timer\! :\! \emph{0},  pedWaiting\! :\! B2 >
   => 
    if americanXing then
      < carLight(CN,\! DIR)\! :\! CarLight | state\! :\! toGreen, 
                                      timer\! :\! Delta + yellowTime + safetyMargin > 
      < approach(CN,\! DIR)\! :\! XingApproach | >
    else --- european, timer only to Delta and another state:
      < carLight(CN,\! DIR)\! :\! CarLight | state\! :\! toRedYellow, timer\! :\! Delta + yellowTime >
      < approach(CN,\! DIR)\! :\! XingApproach | > fi
   if (B1 or B2) . 
\end{alltt} \normalsize

In the following rule, the car light turns green and the @timer@ is set so that
the light  stays green for time @GT@
(which is the variable denoting the value of the attribute
@greenTime@). If a pedestrian is recorded to be waiting, a @pedGo@
message is also sent to the corresponding pedestrian light:

\small \begin{alltt}
rl [redToGreen] :
   < CL\! :\! CarLight | state\! :\! toGreen, greenTime\! :\! GT, timer\! :\! 0, pedWaiting\! :\! B >
  =>
   < CL\! :\! CarLight | state\! :\! green, timer\! :\! GT, lights\! :\! green, pedWaiting\! :\! false >
   (if B then pedGo(pl(CL), GT) else none fi) .
\end{alltt} \normalsize

When the car light is 
in state @green@ and the timer has expired, the light should turn yellow
and stay yellow for time @yellowTime@. Then, it should turn red:

\small \begin{alltt}
rl [greenToYellow] :
   < CL\! :\! CarLight | state\! :\! green, timer\! :\! 0 >
  =>
   < CL\! :\! CarLight | state\! :\! yellow, timer\! :\! yellowTime, lights\! :\! yellow > .

rl [goRed] :
   < CL\! :\! CarLight | state\! :\! yellow, timer\! :\! 0, redTime\! :\! RT >
  =>
   < CL\! :\! CarLight | state\! :\! red, timer\! :\! RT monus (Delta + yellowTime + safetyMargin),
                   \, lights\! :\! red > .
\end{alltt} \normalsize

The following two rules deal with receiving messages from the
pedestrian light when someone wants to cross the intersection. If the
car light in the same direction as the pedestrian crossing
is green, and there is enough time for the pedestrian to
cross during the car light's remaining time in green
 (@TI >= walkTime@), then the car light sends a @pedGo@ signal to the
pedestrian light with the time remaining of the green light (rule 
@buttonPressedTurnedOn@). If the car light in the same direction as
the pedestrian crossing does not show green or if
there is not enough time remaining for the green light to allow the
pedestrian to cross safely, the car light remembers that a
pedestrian is waiting by setting @pedWaiting@ to @true@:

\small \begin{alltt}
crl [buttonPressedTurnOn] :
    (to CL pedsWaiting)   
    < CL\! :\! CarLight | state\! :\! green, timer\! :\! TI > 
   => 
    < CL\! :\! CarLight | >    
    pedGo(pl(CL), TI)
   if TI >= walkTime .
                  
crl [rememberButtonPressed] :
    (to CL pedsWaiting)   
    < CL\! :\! CarLight | state\! :\! S, timer\! :\! TI >
   =>
    < CL\! :\! CarLight | pedWaiting\! :\! true >
   if (S =/= green) or (TI < walkTime) .
\end{alltt} \normalsize

Finally, if the car light receives a @continueGreen@ message, it knows
that it can stay green for another round, and increases its timer by a
whole round:

\small \begin{alltt}
rl [continueGreen]\! :\! 
   continueGreen(CL)
   < CL\! :\! CarLight\! |\! state\!\! :\! NORMAL, timer\!\! :\! T, greenTime\!\! :\! GT, redTime\!\! :\! RT, pedWaiting\!\! :\! B\!\! >
  =>
   < CL\! :\! CarLight\! |\! timer\! :\! T + GT + RT + yellowTime, pedWaiting\! :\! false >
   (if B then pedGo(pl(CL), T + GT + RT + yellowTime) else none fi) .
\end{alltt} \normalsize

\subsubsection{Pedestrian Lights During Normal Operations}

Our pedestrian lights work by sending @pedWaiting@ messages to the car
light when a button is pushed and the pedestrian cannot cross, by
receiving @pedGo@ messages from the car light, which turns the pedestrian light
green, and by starting blinking and then turning red at appropriate
times thereafter.

In the following declaration of the class @PedLight@, the attribute
@color@ shows the color of the light, where we have now added 
@blinking@ as a ``color''; @buttonLit@ is true if some pedestrian has
pushed the button and has not been able to cross since; and @mode@
denotes the mode the system is assumed to be in, which is @normal@
during normal operations:

\small  \begin{alltt}
  class PedLight | timer\! :\! TimeInf, color\! :\! Color, buttonLit\! :\! Bool, mode\! :\! PLMode . 

  sort PLMode .   ops normal emergency\! :\! -> PLMode [ctor] .
\end{alltt}  \normalsize

In rule @turnGreen@, the pedestrian light receives a @pedGo@ message
with a time @T@ during which it can be green or blinking. It sets the
timer to @T monus walkTime@ (where $x\;@monus@\;y=\max(x-y, 0)$) and
turns the pedestrian light green. When it is time to start blinking,
the rule @startBlinking@ turns the pedestrian light to @blinking@ and
sets the timer to @walkTime@. When the blinking time is over, the
pedestrian light turns red and turns off the timer. The pedestrian
light therefore stays red until it receives a @pedGo@ message:

\small  \begin{alltt}
rl [turnGreen]\! :\! 
   pedGo(PL, T)     
   < PL\! :\! PedLight | mode\! :\! normal >
  =>
   < PL\! :\! PedLight | timer\! :\! (T monus walkTime), color\! :\! green, buttonLit\! :\! false > .

rl [startBlinking] :
   < PL\! :\! PedLight | timer\! :\! 0, color\! :\! green >
  =>
   < PL\! :\! PedLight | timer\! :\! walkTime, color\! :\! blinking > .

rl [stop] :
   < PL\! :\! PedLight | timer\! :\! 0, color\! :\! blinking >
  =>
   < PL\! :\! PedLight | timer\! :\! INF, color\! :\! red > .
\end{alltt}  \normalsize

The following  rules treat the arrival of
new pedestrians. If the light
is blinking or red (@C =/= green@), the pedestrian does not cross; if
in addition the button is not lit, the pedestrian pushes the button,
with the result that the button is lit and a @pedsWaiting@ message is
sent to the corresponding car light (rule @newPedestrian1@). If the
light is green when the pedestrian arrives, (s)he just crosses the
street, and if the button is lit, (s)he just joins the other waiting
pedestrians. In neither of these cases does the new pedestrian cause
any change in the resulting state (rule @newPedestrian2@):

\small  \begin{alltt}
crl [newPedestrian1] :
    newPed(pedStop(CN, DIR))
    < pedLight(CN, DIR)\! :\! PedLight | buttonLit\! :\! false, color\! :\! C >
   =>
    < pedLight(CN, DIR)\! :\! PedLight | buttonLit\! :\! true > 
    (to carLight(CN, DIR) pedsWaiting)
   if (C =/= green) .

crl [newPedestrian2] :
    newPed(pedStop(CN, DIR))
    < pedLight(CN, DIR)\! :\! PedLight | buttonLit\! :\! B, color\! :\! C >
   =>
    < pedLight(CN, DIR)\! :\! PedLight | > 
   if (C == green) or B .
\end{alltt}  \normalsize

\subsubsection{Cars}

Most intersections  have sensors that sense if a
car is close to the intersection. For each direction, we model this
sensor as an object of the class

\small  \begin{alltt}
  class XingApproach | carsPresent\! :\! Bool .
\end{alltt}  \normalsize

\noindent whose @carsPresent@ attribute is @true@ if one or more cars are
present close to the intersection. The treatment of new cars from the
environment is  thus modeled by the following rule:

\small  \begin{alltt}
rl [newCars] :
   newCars(CA)   
   < CA\! :\! XingApproach | >
  =>
   < CA\! :\! XingApproach | carsPresent\! :\! true > .
\end{alltt}  \normalsize

In addition, when the car light is green, \emph{all} cars at the
intersection \emph{may} be able to cross the intersection. The
following rule is nondeterministic in the sense that it is not triggered by the
arrival of a message or by the expiration of a timer. It therefore may
or may not be applied when  enabled:

\small  \begin{alltt}
rl [allCarsPass] :
   < approach(CN, DIR)\! :\! XingApproach | carsPresent\! :\! true >
   < carLight(CN, DIR)\! :\! CarLight | lights\! :\! green >
  => 
   < approach(CN, DIR)\! :\! XingApproach | carsPresent\! :\! false >
   < carLight(CN, DIR)\! :\! CarLight | > .
\end{alltt}  \normalsize

\subsubsection{Emergency Handling}

The system handles an emergency by turning all
lights red, after first turning a green light yellow to avoid further
accidents. After the end of the emergency, the lights in the prioritized direction turn green, 
provided that some car or pedestrian
wants to go in that direction. 
We present below two of the eight rules that model the
emergency-related behavior of the car lights. 

The following rule handles the arrival of an emergency signal when the
car light is in a normal (i.e., non-emergency and non-failure)
internal state. If the light is green, then it must first  turn yellow before
turning red after time @yellowTime@. However, if the light is already
yellow, the remaining time that it stays yellow doesn't change. In
addition, an @emergency@ message is sent to the corresponding
pedestrian light, and the 
car light controller goes to internal state @emergency@:

\small  \begin{alltt}
rl [newEmergency] :
   emergency(CL)
   < CL\! :\! CarLight | state\! :\! NORMAL, timer\! :\! T >
  =>
   < CL\! :\! CarLight | state\! :\! emergency, 
                    \,timer\! :\! (if (NORMAL == green) then yellowTime
                             else (if NORMAL == yellow then T else INF fi) fi),
                    \,lights\! :\! (if NORMAL == green or NORMAL == yellow
                              then yellow else red fi) >
   emergency(pl(CL)) .      --- send emergency message to ped light
\end{alltt}  \normalsize

The following rule models the arrival of an @emergencyOver@
message. The car light in the prioritized direction 
 (@defaultStarter : true@) is supposed to turn green, \emph{provided}
that 
cars are present in its direction in the intersection or that it has
recorded that pedestrians are waiting (@B1 or B2@). The car light
turns green, and tells the other car light to restart by showing red
(@reStartRed@). In addition, the car light has to signal to its
pedestrian light that the emergency  is over by sending either a
 @resumeGreen@ message or a @resumeRed@ message, depending on whether
or not pedestrians are waiting in this direction:

\small  \begin{alltt}
crl [emergencyOverMainDirectionStart] :
    emergencyOver(carLight(CN, DIR))
    < approach(CN, DIR)\! :\! XingApproach | carsPresent\! :\! B1 >
    < carLight(CN, DIR)\! :\! CarLight | state\! :\! emergency, defaultStarter\! :\! true,
                                    \,greenTime\! :\! GT, pedWaiting\! :\! B2 >
   =>
    < approach(CN, DIR)\! :\! XingApproach | >
    < carLight(CN, DIR)\! :\! CarLight | state\! :\! green, timer\! :\! GT, 
                                    \,lights\! :\! green, pedWaiting\! :\! false >
    reStartRed(carLight(CN, opposite(DIR)))
    (if B2 then resumeGreen(pedLight(CN,\,DIR),\,GT) else resumeRed(pedLight(CN,\,DIR)) fi)
   if B1 or B2 .
\end{alltt}  \normalsize

\subsubsection{Device Failures}

Any device may fail at any time. Upon a failure, a device must
immediately signal to all the other devices in the crossing, so that the
system begins with the  failure treatment. We choose to deal with failures by
letting the car lights in the prioritized  direction blink yellow, by letting
the car lights in the other direction blink red, and by turning off
the pedestrian lights. The following new ``colors'' are therefore
introduced:

\small  \begin{alltt}
ops blinkingYellow blinkingRed off\! :\! -> Color [ctor] .
\end{alltt}  \normalsize

Since any subset of the four devices in an intersection may fail at
any time, each object must keep track of the number of temporarily 
failed devices. We therefore add a new constructor for keeping track
of errors in the internal state of the controllers:

\small  \begin{alltt}
sort ErrorState .   subsort ErrorState < CLState .
op error\! :\! NzNat -> ErrorState [ctor] .
op errorRecovery\! :\! -> CLState [ctor] .
\end{alltt}  
\normalsize

We present below two of the thirteen rules that define the treatment of
failures and failure recovery. 

The following rule defines the treatment of an @error@ message when
the system is \emph{not} already in an error state
 (@not (S :: ErrorState)@; however, the controller may be
in the @emergency@ state). 
The light controller goes into state @error(1)@, and if it is the light in the prioritized direction, 
 it starts blinking yellow, otherwise it
starts blinking red. 
If it is the device itself that has failed
(the argument of the @error@ message denotes the failed device), it must
notify the three other devices about its failure:

\small  \begin{alltt}
crl [somethingBroken1] :
    (to CL error(DID))
    < CL\! :\! CarLight | defaultStarter\! :\! B, state\! :\! S >
   =>
    < CL\! :\! CarLight | lights\! :\! (if B then blinkingYellow else blinkingRed fi),
                      state\! :\! error(1),timer\! :\! INF  >
    --- If broken device was my device: send messages to other devices:
    (if CL == DID then  (to opposite(CL) error(CL))  (to pl(CL) error(CL))
                        (to pl(opposite(CL)) error(CL))
     else none fi)
   if not (S :: ErrorState) .
\end{alltt}  \normalsize

When the last failed device has been repaired, the lights go back to
the normal state. However, to avoid inconsistent situations, in which a
message from the environment may be read just before or after the last
repair, we have to enforce a regime where messages (other than error
and repair messages) are ignored for a time @Delta@ after the last
repair. 

In the rule @lastDeviceFixed@, the car light has recorded that one
device is still broken (@error(1)@), and a @repaired@ message
arrives. The car light goes to the @errorRecovery@ state and stays
there for a short time @Delta@. If the car light is in the prioritized direction
(\texttt{defaultStater} is \texttt{true}), 
 it turns green, otherwise red. (In this case, there is no
check whether or not pedestrians are waiting, since in error mode,
messages from the pedestrian lights are ignored.)  As always, if the
repaired device was this device (@CL == DID@),
 it has to notify the other devices
about the repair:

\small  \begin{alltt}
rl [lastDeviceFixed]\! :\! 
   (to CL repaired(DID))
   < CL\! :\! CarLight | state\! :\! error(1), defaultStarter\! :\! B >
  =>
   < CL\! :\! CarLight | state\! :\! errorRecovery, timer\! :\! Delta,
                   \, lights\! :\! (if B then green else red fi) >
   (if CL == DID then  (to opposite(CL) repaired(CL))   
                       (to pl(CL) repaired(CL))
                       (to pl(opposite(CL)) repaired(CL)) 
    else none fi) .
\end{alltt}  \normalsize

After time @delta@ in @errorRecovery@ mode, the system goes back to
normal mode (not shown).

\subsubsection{Modeling Environments}

The following class @PeriodicEnv@ is used to periodically generate
nondeterministically a subset (which may be empty) of a set of
messages:

\small  \begin{alltt}
class PeriodicEnv | frequency\! :\! NzTime,  timeToNextEvents\! :\! TimeInf,
                    \,possibleEvents\! :\! NEMsgConfiguration .
\end{alltt}  \normalsize

\noindent The @frequency@ attribute states at what interval the environment
object should generate a subset of the set of messages in the
@possibleEvents@ attribute. (If @frequency@ equals 1 and the time
 domain is discrete, then a subset is
 generated at any moment in time.)

The environment generating new cars and pedestrians is defined by the 
following rule, where the multiset \texttt{ MsgSET1 MsgSET2 } is the total
 set of messages that can be
generated. The rule generates the message set
@MsgSET1@ when the timer expires. This set @MsgSET1@ could be \emph{any}
 subset of @MsgSET1 MsgSET2@, since  message multisets  are
defined  with a multiset union operator that is
 associative and commutative and has \texttt{null} as its identity.

\small  \begin{alltt}
var E\! :\! Oid .  vars MsgSET1 MsgSET2\! :\! Configuration .   var NZT\! :\! NzTime .

rl [generateSubsetAndReset] :
   < E\! :\! PeriodicEnv | timeToNextEvents\! :\! \emph{0}, frequency\! :\! NZT, 
                     \, possibleEvents\! :\! MsgSET1 MsgSET2 >
  =>
   < E\! :\! PeriodicEnv | timeToNextEvents\! :\! NZT >
   \emph{MsgSET1} .
\end{alltt}  \normalsize

The generator of @emergency@ and @emergencyOver@ messages is similar,
with the difference being that @emergency@ and @emergencyOver@ messages
should be generated in an alternating way, so that an @emergencyOver@
message is generated when @emergencyOn@ is true. 
The  environment that generates failures and repairs
 of the devices is defined in a similar way, with the difference being that
there should be some minimum time interval between the repair of a device 
 and the 
next failure of the same device. 

\subsubsection{Time Advance Behavior}

The time advance behavior of the system is defined as for most object-oriented
Real-Time Maude specifications (see~\cite{journ-rtm} and the executable Real-Time Maude model),
 and is not further explained here. 
The tick rule definition ensures that time advance stops when a timer expires, and that
time does not advance if there are messages in the state. This 
forces all messages to be read without delay.

\section{Analyzing the Model}
\label{sec:analysis}

This section shows how our model can be analyzed by using 
\emph{linear
temporal logic} (LTL), including \emph{metric} LTL,  model checking to analyze all possible behaviors
from a given initial state.

\subsection{Defining Initial States}
\label{sec:initial-states}

In the spirit of the SoW, we have defined initial states that are
parametric in the duration of the lights, the number of
flawed devices, the possibility of  emergencies,  the
duration between possible failures, etc. 

First, we define the light objects in an intersection. The term
@lights@(\emph{name}, \emph{dir}, \emph{greenTime}, \emph{redTime})
defines the car light and pedestrian light objects for an intersection
called \emph{name}, where  direction \emph{dir} is the prioritized direction,
 and where the green time and red time of this light
is, respectively,   \emph{greenTime} and \emph{redTime}:

\small  \begin{alltt}
eq lights(CN, DIR, GREENTIME, REDTIME) =
   < carLight(CN,\! DIR)\! :\! CarLight | lights\! :\! green, timer\! :\! GREENTIME, redTime\! :\! REDTIME, 
                                   greenTime\! :\! GREENTIME, state\! :\! green, 
                                   pedWaiting\! :\! false, defaultStarter\! :\! true >
   < carLight(CN,\! opposite(DIR))\! :\! CarLight | lights\! :\! red, 
                                             timer\! :\! (GREENTIME monus Delta),
                                             redTime\! :\! (GREENTIME + yellowTime +
                                                    \!    safetyMargin + safetyMargin),  
                                             greenTime\! :\! (REDTIME monus (yellowTime + 
                                                        \!\!  safetyMargin + safetyMargin))\!,
                                             state\! :\! red, pedWaiting\! :\! false, 
                                             defaultStarter\! :\! false >
   <\! pedLight(CN,\! DIR)\!\! :\!\! PedLight\! |\! timer\!\! :\! INF\!, color\!\! :\! red, buttonLit\!\! :\! false\!, mode\!\! :\! normal\!\! > 
   < pedLight(CN,\! opposite(DIR))\! :\! PedLight | timer\! :\! INF, color\! :\! red,
                                             \,buttonLit\! :\! false, mode\! :\! normal > 
   < approach(CN,\! NS)\! :\! XingApproach | carsPresent\! :\! false >
   < approach(CN,\! EW)\! :\! XingApproach | carsPresent\! :\! false > .
\end{alltt}  \normalsize  

The environment object that generates subsets of cars
and pedestrians every @NZT@th time unit in a crossing  @CN@ is defined as
follows:

\small  \begin{alltt}
op carsAndPeds\! :\! CrossingName NzTime -> Configuration .
eq carsAndPeds(CN,\! NZT) =
     < carsAndPeds(CN)\! :\! PeriodicEnv | frequency\! :\! NZT, timeToNextEvents\! :\! 0,
                                       possibleEvents : (newCars(approach(CN,\! NS))
                                                         newCars(approach(CN,\! EW))
                                                         newPed(pedStop(CN,\! NS))
                                                         newPed(pedStop(CN,\! EW))) > .
\end{alltt}  \normalsize 

Emergency generators and error generators can be defined  in the same way.

A  term \texttt{init(XING, GREENTIME, REDTIME, T, CARF, PEDF, N1, N2)} defines an initial state, 
 with an intersection called \texttt{XING}, where 
 NS is the prioritized direction, and where the NS light
has green time \texttt{GREENTIME} and red time \texttt{REDTIME}. The time
between each nondeterministic ``choice'' of whether or not to generate
an @emergency@ or @emergencyOver@ message is denoted by 
\texttt{T} (if this number  is 0, then no emergencies
 are generated). The
number of potentially faulty car lights and pedestrian lights is
denoted by, respectively,  \texttt{CARF}  and \texttt{PEDF}. Finally, \texttt{N1}
denotes how often errors may occur, and \texttt{N2} denotes the
minimum time between the repair of a device and the next failure of
that device:

\small  \begin{alltt}
eq init(CN, GREENTIME, REDTIME, T, CARF, PEDF, NZT, NZT1) =
     \texttt{\char123}lights(CN, NS, GREENTIME, REDTIME)
      carsAndPeds(CN, 1)
      (if T =/= 0 then emergencyEnv(CN, T) else none fi)
      (if CARF == 2 then bothCarErrors(CN, NZT, NZT1)
       else (if CARF == 1 then  errors(carLight(CN, NS), NZT, NZT1) else none fi) fi)
      (if PEDF == 2 then errors(pedLight(CN, NS), NZT, NZT1)
                         errors(pedLight(CN, EW), NZT, NZT1) 
       else (if PEDF == 1 then errors(pedLight(CN,\! EW),\! NZT,\! NZT1) else none fi) fi)\texttt{\char125}\! .
\end{alltt}  \normalsize

\subsection{LTL and Metric LTL Model Checking}

We can now use Real-Time Maude's LTL model checker to analyze
 our model with respect to the requirements
given in the SoW.

The first property to check is requirement (v) (``the system shall turn green a 
pedestrian light only when there are pedestrians waiting to cross in that direction''). 
In our model, the  fact that a pedestrian is waiting is  represented
 by the fact that the button on the pedestrian light pole is ``lit.''
 The LTL formula in the following model checking command states that,
 for \emph{all} states reachable from the initial state\footnote{Remember that
$\phi\; @=>@\; \theta$ is an abbreviation for
$@[]@ (\phi \; @->@\; \theta)$.}, and for both
 directions NS and EW, if the pedestrian light is red and the button is not
 lit (@(pedLightRed(EW) /\ ~ buttonPushed(EW))@), then the  
  light remains red  forever, or 
 until a pedestrian arrives 
(@(pedLightRed(EW) W pedArriving(EW))@). 
   This property is checked by the following Real-Time Maude command
 in the presence of 
 arbitrary emergency signals and with \emph{no} failed devices:

\small  
\begin{verbatim}
(mc init("Spitsbergen", minGreenTime + 2, minRedTime, 2, 0, 0, 1, 1) 
       |=u
    ((pedLightRed(EW) /\ ~ buttonPushed(EW)) => (pedLightRed(EW) W pedArriving(EW)))
        /\
    ((pedLightRed(NS) /\ ~ buttonPushed(NS)) => (pedLightRed(NS) W pedArriving(NS))) .)
\end{verbatim}  
\normalsize

The proposition @pedLightRed(@$d$@)@ holds if the pedestrian light in
direction $d$ shows red:

\small  
\begin{alltt}
op pedLightRed : Direction -> Prop [ctor] .
eq \char123REST  <\! pedLight(O,\! DIR)\! :\! PedLight\! |\! color\! :\! C\! >\char125 |= pedLightRed(DIR) = (C == red) .
\end{alltt}  \normalsize 

\noindent  The proposition @pedArriving@ is @true@ if
the state contains a message @newPed@ for the given direction:

\small  
\begin{alltt}
op pedArriving\! :\! Direction -> Prop [ctor] .
eq \char123REST  newPed(pedStop("Spitsbergen", DIR))\char125  |=  pedArriving(DIR) = true .
\end{alltt}  \normalsize 

\noindent The  proposition @buttonPushed@ can be defined in a similar way. 
The model checking command above returned the expected result @true@ after
executing for 212 seconds on a server.

The next property we check is a modification of requirement (vi) in the SoW, namely, that
 the system should turn a vehicular light green only if
 there are vehicles \emph{or pedestrians} waiting to go in that direction.
 The following formula,  again checked in the presence
 of emergencies, says that if the NS car light is red, and no
 pedestrians or cars are waiting, then this light will be red until
 a pedestrian or a car arrives:

\small  
\begin{verbatim}
(mc init("Spitsbergen", minGreenTime + 2, minRedTime, 2, 0, 0, 1, 1) 
       |=u
    (carLightRed(NS) /\ ~ buttonPushed(NS) /\ ~ carWaiting(NS))
    =>  (carLightRed(NS) W (pedArriving(NS) \/ carArriving(NS))) .)
\end{verbatim}  \normalsize  

We refer to the Real-Time Maude specification for the
definition of the propositions in this formula. 
The result of executing the command above is @true@, and the execution
took 153 seconds. 

Another requirement of the system is that 
    \emph{"The system shall be fault-tolerant."}
 We have defined a set of properties, including some
 of the properties described below, which together
 demonstrate that the system is fault-tolerant. The specific
 liveness property we map
 to this item says that if each failure is eventually repaired
 (@[] (failure -> <> repair)@), 
 then cars in the  NS direction will be able to pass
 infinitely often (@[] <> carLightGreen(NS)@)
 if cars arrive in this direction infinitely often. This 
 property is checked with one faulty device and where
 the smallest time interval between a repair and another failure
 is 9:

\small  
\begin{alltt}
(mc init("Spitsbergen", minGreenTime + 2, minRedTime, 0, 1, 0, 2, 9)
       |=u
    ([] (failure -> <> repair))\!  
    ->\!  (([] <> carArriving(NS)) -> ([] <> carLightGreen(NS)))\! .)
\end{alltt}  \normalsize 

\noindent The propositions @failure@ and @repair@ are true in a state if an
@error@, respectively @repaired@, message appears in the state:

\small  
\begin{alltt}
ops failure repair : -> Prop [ctor] .
eq \char123REST  (to DID error(DID))\char125  |=  failure = true .
eq \char123REST  (to DID repaired(DID))\char125  |=  repair = true .
\end{alltt}  
\normalsize

\noindent The execution of this command returned @true@ in 148  seconds.

The SoW requires that \emph{
    "The system shall be fail-safe, ensuring both failure recovery, and safe
 car and pedestrian conditions also under failure."}
We check whether cars can collide with pedestrians
 in the presence of one faulty device (and no emergencies):

\small  
\begin{verbatim}
(mc init("Spitsbergen", minGreenTime + 2, minRedTime, 0, 1, 0, 2, 9) 
       |=u
    [] ((~ (walking(NS) /\ driving(EW))) /\ (~ (walking(EW) /\ driving(NS)))) .)
\end{verbatim}  \normalsize

The propositions @walking@ and @driving@ are again defined in a
straight-forward way:

\small  
\begin{verbatim}
ops walking driving : Direction -> Prop [ctor] .
eq {REST  < pedLight(O, DIR) : PedLight | color : C >}
     |=
   walking(DIR) = (C == green or C == blinking) .

eq {REST  < carLight(O, DIR) : CarLight | lights : green >} |=  driving(DIR) = true .
\end{verbatim}  \normalsize

\noindent Model checking this crucial safety property resulted in the answer
@true@. We have also checked the  crucial property that it is not possible for cars 
to legally drive in 
both  directions at the same time. 

One QoS requirement  in the SoW  states
that no pedestrian should wait for more than five minutes.
This requirement is analyzed  with the following 
\emph{bounded response} command, that states that each time a pedestrian arrives
in the NS direction, (s)he will be walking within \emph{15 time units},
 in a setting without failures or emergencies
(since there is no upper limit on the duration of an emergency or 
a failure of a device):

\small
\begin{alltt}
(br init("Spitsbergen", minGreenTime + 2, minRedTime, 0, 0, 0, 1, 1, false, 0) 
       |= 
    \,pedArriving(NS) => <>le( 15 ) walking(NS) .)
\end{alltt}
\normalsize

\noindent This command returned @true@ in 168 seconds. Executing the  command
 for 14 time units returned
 a counterexample, so for the given parameter values, 15 time units is the best
  time guarantee we can give.

%% file: concl.tex
\section{Conclusions}

We have presented a highly decentralized and modular parametrized
Real-Time Maude model of a four-way traffic intersection, in which
autonomous devices communicate by asynchronous message passing without
a centralized controller.  All the safety requirements and a liveness
requirement informally speciﬁed in the SoW requirements document have
been formally verified.  We believe that this work shows how formal
specification and verification can be inserted within a design
environment for DES product families.  And that this can be done at
the stage where this matters most, namely, at the design stage, since
design errors are much more costly than coding errors. In particular,
our modeling and formal analysis have allowed us to both obtain a
fully verified model with respect to the required safety properties,
and to identify nontrivial inconsistencies in the requirements
document. 

This positive experience should not be exaggerated and should be
treated with caution.  Scalability is still a very serious challenge
for DES verification.  We feel that in this case study we were at the
limit of the kind of distributed system complexity that can be
\emph{directly} verified with Real-Time Maude, which uses a
state-of-the-art explicit-state model checker.  However, some recent
methods look promising to  
verify properties of more complex DES system designs either indirectly
or under statistical guarantees, including: (i) more aggressive uses of abstraction
methods such as the PALS methodology \cite{dasc09,pals-techrep,pals-rtss09,meseguer-olveczky-pals-icfem,meseguer-olveczky-pals-rep}, which can greatly decrease system
complexity and verification cost by reducing the verification of
asynchronous real-time systems to that 
of a much simpler synchronous version under some conditions; and (ii) the use of
more scalable \emph{statistical model checking}~\cite{younesS06,SVAcav05}
 techniques that trade off full verification
of system properties by statistical guarantees on the satisfaction of such properties
when full model checking verification becomes unfeasible.  

\small
\textbf{Acknowledgments.} 
This work is part of the Lockheed Martin Advanced Technology
Laboratories' NAOMI project. We thank  Edward Jones and Trip Denton
of Lockheed Martin for providing the informal specification and requirements
for the case study, and the anonymous reviewers 
for very helpful comments on an earlier version of this paper.
 We also gratefully acknowledge financial support by
Lockheed Martin, through the NAOMI
project, the NSF, through Grant CNS 08-34709,
and  the Research Council of Norway, through the Rhytm project.
\normalsize